\documentclass[aps,prb,floatfix,showpacs,preprintnumbers,amsmath,amssymb]{revtex4}
\usepackage{graphicx}% Include figure files
\usepackage{dcolumn}% Align table columns on decimal point
\usepackage{bm}% bold math

\begin{document}

\title{Electronic Structure of Transition-Metal Dicyanamides \\
  Me[N(CN)$_2$]$_2$ (Me = Mn, Fe, Co, Ni, Cu)}

\author{D. O. Demchenko and Amy Y. Liu}
\affiliation{Department of Physics, Georgetown University, Washington, DC 
20057, USA}
\author{E. Z. Kurmaev, L. D. Finkelstein and  V. R. Galakhov}
\affiliation{Institute of Metal Physics, Russian Academy of Sciences-Ural 
Division, 620219 Yekaterinburg GSP-170, Russia}
\author{A. Moewes}
\affiliation{Department of Physics and Engineering Physics, University of 
Saskatchewan, 116 Science Place, Saskatoon, Saskatchewan S7N 5E2, Canada}
\author{S. G. Chiuzb\u{a}ian\footnote{Present address: Paul Scherrer Institut, 
Swiss Light Source, CH-5232 Villigen, Switzerland} and M. Neumann}
\affiliation{Universit\"at Osnabruck, Fachbereich Physik, D-49069 Osnabruck, 
Germany}
\author{Carmen R. Kmety}
\affiliation{Advanced Photon Source, Argonne National Laboratory, 
Argonne IL 60439, USA}
\author{Kenneth L. Stevenson}
\affiliation{Department of Chemistry, Indiana University Purdue University, 
Fort Wayne, Indiana 46805, USA}
\date{\today}

\begin{abstract}
The electronic structure of Me[N(CN)$_2$]$_2$ 
(Me=Mn, Fe, Co, Ni, Cu) molecular magnets has been investigated using  
x-ray emission spectroscopy (XES) and x-ray photoelectron 
spectroscopy (XPS) as well as theoretical density-functional-based
methods.  Both theory and experiments show that the top of the valence band
is dominated by Me $3d$ bands, while a strong 
hybridization between C $2p$ and  N $2p$ states determines the 
valence band electronic structure away from the top. 
The $2p$ contributions 
from non-equivalent nitrogen sites have been identified using 
resonant inelastic x-ray scattering spectroscopy  with the 
excitation energy tuned near the N $1s$ threshold. 
The binding energy of the Me $3d$ bands and the   
hybridization between N $2p$ and Me $3d$ states both increase 
in going across the row from Me = Mn to  Me = Cu.  
Localization of the Cu $3d$ states also leads to weak screening
of Cu $2p$ and $3s$ states, which accounts for shifts in the core $2p$ and $3s$
spectra of the transition metal atoms. 
Calculations indicate that the ground-state magnetic ordering, 
which varies across the series is 
largely dependent on the occupation of the metal $3d$ shell and that 
structural differences in the superexchange pathways for different 
compounds play a secondary role. 
\end{abstract}


\maketitle

\section{Introduction}

Molecule-based magnets have been a subject of interest in the past
decade, after the discovery of room-temperature magnetic ordering in
some of these materials.\cite{Manriquez, Ferlay}  Interest is driven not 
only by the potential for use in applications like information storage and 
processing, but also by fundamental questions about the origin of
magnetic ordering in these materials.   The transition-metal
dicyanamides, Me[N(CN)$_2$]$_2$ where Me is a $3d$ transition metal,
provide an example where, even within an isostructural
series of compounds,  the magnetic ordering can differ substantially 
depending on the metal species and details of the crystal structure. 
Experimental studies of the structural and magnetic properties of the 
dicyanamides with Me = Mn, Fe, Co, Ni, and Cu have been performed by 
several groups.\cite{Kmety3, Batten, Kurmoo, Kmety1, Kmety2, thesis}  
Mn[N(CN)$_2$]$_2$ and Fe[N(CN)$_2$]$_2$ have been found to be 
canted antiferromagnets, while Co[N(CN)$_2$]$_2$ and Ni[N(CN)$_2$]$_2$
have been found to be ferromagnets. Cu[N(CN)$_2$]$_2$ has been reported 
to be paramagnetic,
at least down to a few Kelvin, and a recent experiment has found that 
it becomes a ferromagnet below 1.8 K. \cite{thesis}
Thus the compounds with transition-metal ions having six or fewer
$3d$ electrons order as canted antiferromagnets, while those with seven
or more $3d$ electrons order as ferromagnets.  The Curie (or N\'{e}el)
temperatures of these materials range from 1.8 K for Cu[N(CN)$_2$]$_2$
to 22.7 K for Ni[N(CN)$_2$]$_2$.  Ni[N(CN)$_2$]$_2$ and Fe[N(CN)$_2$]$_2$ 
have been found to have  the largest coercive fields of all known 
metal-organic magnets -- 7975 and 17800 Oe, respectively.

The molecular building block of the bulk dicyanamides consists of
a divalent metal ion that is two-fold coordinated by  two N(CN)$_2$ groups.
When the molecular units pack to form the bulk crystal,  
the negatively charged CN groups at the end of neighboring molecules
interact with the positive central metal ion in such a way that the metal
ion becomes surrounded by an octahedron of six N atoms.  The
local symmetry around the metal site is tetragonal, with the
original molecular axis corresponding to the long axis.   The 
primitive cell contains two formula units with their long axes rotated in
opposite directions in the $xy$ plane, leading to an orthorhombic cell.
The metal ions are the source of magnetic moments while the
organic species  provide superexchange pathways between
the magnetic centers.  

In this paper, we report a combined experimental and theoretical
study of the electronic structure of Me[N(CN)$_2$]$_2$ with Me = Mn,
Fe, Co, Ni, Cu.   We have measured x-ray emission spectra (XES)
and x-ray photoemission spectra (XPS) across the series. 
Generally, XPS probes the total occupied density of
states (DOS) whereas XES probes the site-selective partial DOS due to the
dipole selection rules. In the case of carbon and nitrogen $K_\alpha$  XES,
which correspond to $2p \rightarrow 1s$ transitions, the occupied $2p$ states
can be studied. Similarly, Me $L_{2,3}$ XES, related to the
$3d4s \rightarrow 2p$ transition, gives information about occupied $3d$ bands.
The comparison of XPS and XES in the binding-energy scale yields information
about the distribution of the partial and total DOS in multicomponent systems.
The measured spectra are compared with results of density-functional 
calculations.  In a recent paper focusing on the 
electronic structure of Mn[N(CN)$_2$]$_2$, reasonable agreement was found 
between theory and experiment for the electronic structure, magnetic ordering, 
and magnetic anisotropy.\cite{Pederson}  Here we find that the direct
comparison of 
density-functional calculations with experimental data
is not as straightforward for some of the other 
compounds in the series.

\section{Experimental and Calculation Details}

The x-ray fluorescence measurements were performed at Beamline 8.0 at the 
Advanced Light Source (ALS) of Lawrence Berkeley National Laboratory (LBNL). 
Carbon and nitrogen $K_\alpha$ ($2p \rightarrow 1s$ transition) and Me 
$L_{2,3}$ ($3d4s \rightarrow 2p$ transition) x-ray emission spectra were 
taken, employing the University of Tennessee at Knoxville soft x-ray 
fluorescence endstation.\cite{Jia} Photons with energy of 300 eV above the 
carbon $K$ edge were delivered to the endstation via the beamline's 89-period, 
5 centimeter undulator insertion device and spherical monochromator. 
For nitrogen $K_\alpha$ and Me $L_{2,3}$ XES, an excitation-energy dependence 
was measured near the N $1s$ and Me $2p$ thresholds. The carbon and nitrogen 
$K_\alpha$ spectra were obtained with an energy resolution of 0.3-0.4 eV. 
The Me $L_{2,3}$ XES were measured with an energy resolution of 0.7-0.8 eV.

The XPS measurements were performed with an ESCA spectrometer of Physical 
Electronics (PHI 5600 ci) with monochromatized Al $K_\alpha$ radiation of a 
0.3 eV FWHM. The energy resolution of the analyzer was 1.5\% of the pass 
energy. The pressure in the vacuum chamber during the measurements was below 
$5\times 10^{-9}$ mbar. Prior to XPS measurements, the samples were fractured 
in ultra-high vacuum. All photoemission studies were performed at room 
temperature on freshly cleaved surfaces. Charging of the insulating 
Me[N(CN)$_2$]$_2$ samples was compensated with an electron gun. 
Because there is no saturated carbon in these molecules, the
binding energies can not be corrected using the C $1s$ line. 
Therefore we have used an alternative way\cite{Moulder} of fixing the 
N $1s$ line of the cyanide group at 399.6 eV.

Polycrystalline pressed pellets of Me[N(CN)$_2$]$_2$ compounds were used for 
XES and XPS measurements. Details about the sample preparation can be found in 
Refs. \onlinecite{Kmety1} and \onlinecite{Kmety2}.

Diffuse reflectance spectroscopy on powder samples was used to measure the 
lowest electronic transition observable in the visible region, which 
can be interpreted as possible band gap energies.  
The reflectance spectra 
were taken on powdered samples at room temperature in a 1-mm quartz cuvette which was illuminated by 
light from a 75-watt Xe high-pressure arc lamp.  The reflected light was admitted into 
an Acton Spectropro 300i 300 mm spectrometer with a Hammamatsu CCD detector, 
which gave light intensities versus wavelength from 300 to 800 nm.
The spectrum of a ``white'' standard, consisting of powdered NaCl, 
was used for determination of the absorbance, $\log(I_0/I_s)$, as a function of wavelength
for each sample, where $I_0$ is the intensity of 
the ``white'' standard and $I_s$ is the intensity of the sample compound.

For comparison with the experimental results, 
electronic-structure calculations were performed for ferromagnetic (FM), 
antiferromagnetic (AF), and nonmagnetic (NM) phases using the crystal 
structures determined from x-ray and neutron diffraction 
experiments.\cite{Kurmoo, Kmety1, Kmety2, Kmety3, thesis} 
Spins were restricted to be collinear. Calculations were carried out using the 
VASP program,\cite{vasp} a density-functional-based code employing planewave 
basis sets and ultrasoft pseudopotentials. The electron-electron interaction 
was treated in the generalized gradient approximation (GGA),\cite{GGA1, GGA2} 
and electronic wave functions were expanded in planewaves up to a kinetic
energy cutoff of 435 eV. Monkhorst-Pack meshes\cite{Monkhorst-Pack}
of up to $8\times 8\times 8$ {\bf k}-points were used to sample the 
Brillouin zone, and the linear tetrahedron method was used to integrate over 
the Brillouin zone.

\section{Magnetic Properties}

%---------------------------------------------------------
Table \ref{summary} summarizes the results of the total-energy calculations.
Density-functional-theory (DFT)  calculations for terminated cluster 
models yield similar results.\cite{Pederson1} 
Experimentally, Mn[N(CN)$_2$]$_2$ and Fe[N(CN)$_2$]$_2$  
are both canted antiferromagnets. In the Mn compound, the canting angle 
is reported to be at most a few degrees,\cite{Kmety1} while in the Fe 
compound, the canting angle is estimated to be 20-23$^\circ$.\cite{thesis} 
Our calculations find Mn[N(CN)$_2$]$_2$
to be antiferromagnetic, but neglect of both the canting and spin-orbit interaction 
in Fe[N(CN)$_2$]$_2$ results in the FM phase being favored. 
Similar results were recently obtained in Ref. \onlinecite{Ruiz}. 
We expect that when spins are not
restricted to be collinear, density functional theory would yield an AF
ground state for Fe[N(CN)$_2$]$_2$.
For Ni[N(CN)$_2$]$_2$ and
Co[N(CN)$_2$]$_2$, we find the ground state to be ferromagnetic, in
agreement with experiments. In the case of Cu[N(CN)$_2$]$_2$, the 
FM and AF phases are calculated to be very close in energy and favored 
over the NM phase.
Magnetization studies show no long-range ordering above 2 K,\cite{Kurmoo}
but a recent study \cite{thesis} finds that Cu[N(CN)$_2$]$_2$ becomes
ferromagnetic at lower temperatures with $T_c \sim 1.8$ K.
Hence we report results of our calculations for the FM phase.

%----new superexchange
In superexchange systems, the magnitude and sign of the one-electron
contribution to the exchange interaction depend on the filling of those orbitals
on neighboring magnetic centers that interact via a bridging ligand. \cite{Anderson, Goodenough, Gudel} 
The relevant orbitals, {\it i.e.}, those with non-zero interaction matrix elements, 
depend in turn on the geometry of the superexchange pathway.  
The magnetic ordering is thus determined by an interplay between the 
occupation of the $d$ shell and the superexchange angle. 
In Me[N(CN)$_2$]$_2$, neighboring Me sites are connected
through a Me-[N-C-N]-Me pathway, where the two N sites along the path
are inequivalent.  The Me-[N-C-N]-Me angle increases monotonically in
going from Mn[N(CN)$_2$]$_2$ to Ni[N(CN)$_2$]$_2$, with a total change of about 2.5 degrees. \cite{Kmety1}
It has been suggested that this structural change in the pathway is responsible
for the different magnetic ground states in the dicyanamides, with the
angle passing through a critical value in going across the $3d$ series. \cite{Kmety1}
Our calculations find that such small changes in structure do not change
the energetic ordering of the magnetic phases. 
The occupation of the $3d$ shell is evidently a more significant factor in determining the
sign of the superexchange interaction in these systems.
%----

Density functional theory typically underestimates magnetic moments, but,
as shown in Table \ref{summary},  discrepancies for Me = Fe and Ni appear 
unusually large. This is likely because the measured magnetic moments in the
table, determined from neutron powder diffraction studies,\cite{Kmety1,thesis} 
include both spin and orbital contributions. For octahedrally
coordinated Me$^{2+}$ cations without spin-orbit coupling, the orbital
moment is expected to be completely quenched for Me = Mn, Ni, and Cu, and
unquenched or partially quenched for Me = Fe and Co.\cite{Figgis}
With spin-orbit coupling, the quenching due to the crystal field is
incomplete, resulting in a Land\'{e} g-factor that can deviate from 
2.\cite{thesis,Kmety2}  With no mechanism for unquenching the orbital moment 
within our calculations, a direct comparison between the theoretical
and experimental moments in Table \ref{summary} is not appropriate.

It is well known that the local spin magnetic moment is related to 
spectral splitting of the $3s$ core-level 
x-ray photoemission spectra in transition metals and their compounds.
The splitting originates from the exchange coupling between the $3s$ hole and the 
$3d$ electrons and the magnitude of the splitting is proportional to 
$(2S+1)$, where $S$ is the local spin of the $3d$ electrons in the 
ground state. \cite{VanVleck} 
The $3s$ spectra of the Me[N(CN)$_2$]$_2$ materials together with those 
of simple oxides MnO, FeO, CoO, NiO, and CuO are shown in Fig. \ref{XPS_3s}. 
Peaks A and B denote $3s^{-1}3d^{n+1}L^{-1}$ final-state configurations, 
and C and D denote $3s^{-1}3d^{n}$ final-state configurations, where 
$L^{-1}$ denotes a ligand hole due to the charge-transfer process. For 
Mn[N(CN)$_2$]$_2$, fitting curves and their component peaks are shown. 
The Mn $3s$ spectrum shows 
two peaks, C and D, which can be attributed to the $3s^{-1}3d^{5}$ final state. 
The magnitude of the $3s$ exchange splitting in Mn[N(CN)$_2$]$_2$, 6.5 eV, is very similar to that in
MnO, 6.2 eV, \cite{Galakhov2} which indicates the Mn
$3d^{5}$ ground-state configuration for Mn[N(CN)$_2$]$_2$. This correlates well
with calculated and measured spin moments of Mn[N(CN)$_2$]$_2$ presented in 
Table \ref{summary}.
The Me $3s$ spectra of
Ni[N(CN)$_2$]$_2$, Co[N(CN)$_2$]$_2$, and Fe[N(CN)$_2$]$_2$ show a
complex structure 
formed by both $3s^{-1}3d^{n}$ and $3s^{-1}3d^{n+1}L^{-1}$ final-state 
configurations. 
Coexistence of both exchange and charge-transfer effects makes the comparison
of the Me $3s$ splitting with the local magnetic moments difficult. 
Similarities with the spectra of the monoxides suggest that the
Me $3s$ energy splitting is systematically reduced with
increasing Me atomic number,
as expected for transition metal ions with nominal $2^+ $ valence
(Table \ref{summary}).
For Cu[N(CN)$_2$]$_2$, the exchange splitting is expected
for the $3s^13d^9$ configuration only, which is localized in the energy
interval from 130 to 135 eV. Low energy resolution limits our ability to
separate this splitting in the present data, but it has been observed
in copper oxides (see Fig. \ref{XPS_3s} and Ref. \onlinecite{Galakhov}).
%**--------------------------------------------------

\section{Electronic Structure} 

Figure \ref{XES}  shows the measured carbon and nitrogen 
$K_\alpha$ XES of the Me[N(CN)$_2$]$_2$ compounds.  As seen, the carbon 
$K_\alpha$ XES (Fig. \ref{XES}a) are very similar across the series because 
of the 
same local atomic and electronic structure of carbon atoms in cyano groups. 
On the other hand, the nitrogen atoms occupy two non-equivalent sites in the 
crystal structure of Me[N(CN)$_2$]$_2$ compounds, where each Me atom has 
two N(1) neighbors along the long axis and four N(2) neighbors from CN 
anions along the other axes.\cite{Kmety1}   The nitrogen $K_\alpha$ XES of 
Me[N(CN)$_2$]$_2$ (Fig. \ref{XES}b) has contributions from both sites. The N 
$K_\alpha$ spectrum of the reference compound Na$_4$Fe(CN)$_6$(OH$_2$O) 
(Fig. \ref{XES}b) shows N contributions from cyano groups and is quite 
different from the non-resonant nitrogen spectra of Me[N(CN)$_2$]$_2$.
In our XPS measurements, the N $1s$ spectrum is split into two 
main lines separated by approximately 1.6 eV, corresponding to the difference 
in the N $1s$ binding energies of the two non-equivalent N sites 
(Table \ref{bindingenergy}).

Additional measurements were made for resonantly excited nitrogen $K_\alpha$
XES in Mn[N(CN)$_2$]$_2$.   The energy dependence of the spectra
when the excitation energy was tuned near the N $1s$ threshold is shown
in Fig. \ref{RXES}a. 
The intensities of the peaks at 389.9, 392.7 and 394.8 eV change 
with excitation energy. The non-resonant spectrum $d$ excited far from 
the N $1s$ threshold can be described as a superposition of spectra 
$b$ and $c$ (Fig. \ref{RXES}b), corresponding to contributions 
from selectively excited non-equivalent nitrogen atoms. Based on our 
measurements of the non-resonant N $K_\alpha$ XES of Na$_4$Fe(CN)$_6$(OH$_2$O) 
(Fig. \ref{XES}b), we can attribute the origin of the resonant spectrum of 
Mn[N(CN)$_2$]$_2$ excited near peaks $a$ and $b$ of the N $1s$ 
TEY to N(2) atoms 
located in cyano groups, whereas the spectrum excited near peak $c$ of the
N $1s$ TEY can be explained by contributions from 
both N(1) and N(2) atoms.

The valence band (VB) XPS and the XES of the constituents of Me[N(CN)$_2$]$_2$ 
are compared in Figs. \ref{Mn_XPS}-\ref{Cu_XPS}. 
We have selected for this comparison the non-resonant carbon and nitrogen 
$K_\alpha$ XES shown in Fig. \ref{XES} and the 
Me $L_3$ XES ($3d4s \rightarrow 2p_{3/2}$ transition) which were excited 
just above the $L_3$ threshold to exclude the excitation of the Me $L_2$ 
XES ($3d4s \rightarrow 2p_{1/2}$ transition) that overlaps the Me $L_3$ 
XES due to spin-orbit splitting. To convert the XES of the constituents to 
the binding-energy scale, we have subtracted emission energies from the 
XPS binding energies of core levels (Table \ref{bindingenergy}). 
According to this comparison, 
the top of the valence band (a) is derived mainly from Me $3d$ and N $2p$ 
states. Towards the middle of the valence band the Me $3d$, N $2p$ and C $2p$ 
states are strongly hybridized, forming a broad structure (b-d). At the bottom 
of the valence band (e) are mixed carbon and nitrogen $2p$ states.
Atomic-like carbon and nitrogen $2s$ states are located in
region (f), with binding energy around 20-25 eV. 

For comparison, we present theoretical results for  Ni[N(CN)$_2$]$_2$
as an example.   The calculated electronic density of states (DOS) of
Ni[N(CN)$_2$]$_2$ is shown in Fig. \ref{dos_Ni}. 
The total DOS is in the lowest panel,
and site- and $l$-projected DOS are plotted in the other panels.  In the lowest
panel, a Gaussian broadening of 1.0 eV has been introduced to simulate the
effect of finite instrumental resolution. The partial DOS in the upper three
panels have been broadened by 0.2 eV to help in identifying features seen in
the experimental XES spectra. 
Features in the DOS are labeled (a)-(f) as in the experimental XPS spectra. 
The nature of the states at different energies is generally
consistent with what is found in the XPS/XES comparison.
Regions (a) and (b) near the top of the valence band are dominated
by Ni $3d$ states, with some hybridization with C and N $2p$ states, and
they correspond to the peak in the Ni $L_3$ XES spectrum.
Regions (c) and (d) contain strong contributions from C $2p$ and N $2p$ 
states, as seen in the C and N XES spectra, as well as some Ni $3d$ states. 
Peak (e), approximately 12 eV below the top of the valence band, 
is primarily of C $2p$ and C $2s$ character, and N $2p$ contributions 
from N(1) sites.  Region (f) has significant contributions from the 
N $2s$, C $2s$, and C $2p$ states and is observed in the 
C $K_\alpha$ XES spectrum. 

For the other compounds, the electronic DOS is very similar to that
of Ni[N(CN)$_2$]$_2$ from about 25 eV below the top of the valence band up
to region (d) where $3d$ states of the transition metal start playing a 
significant role. The hybridization between N $2p$ and Me $3d$ states 
increases with the Me atomic number. For Me = Mn, Fe, and Co, states near the
top of the valence band are overwhelmingly of Me $d$ character, while for
Me = Ni and Cu, the relative contribution of N $p$ states is increased. 
In both the calculations and the experiments, the weight of the
$3d$-derived bands shifts with respect to the top of the valence band. Based on
the XPS valence bands, the maximum of the $3d$ bands varies from about 4 eV
for Mn[N(CN)$_2$]$_2$ to about 6 eV for Cu[N(CN)$_2$]$_2$ (see Figs. 
\ref{Mn_XPS}-\ref{Cu_XPS} 
and Table \ref{bindingenergy}).  This result is confirmed by the Me $L_3$ XES 
converted to the binding-energy scale. 
In addition,  
in the calculations, we find, for example, about a  1.5 eV difference in 
the position of the main maxima of the occupied $d$-DOS weights in 
Cu[N(CN)$_2$]$_2$ compared to Ni[N(CN)$_2$]$_2$ (Fig. \ref{dos_Cu}), 
which is similar to what is observed experimentally (Table \ref{bindingenergy}).
The deepening of $3d$ states found for Cu[N(CN)$_2$]$_2$ is due 
to the localization $d$ electrons at the end of the $3d$  series. This is 
observed in molecular crystals because the $d-d$ and $d-sp$ interactions 
are weaker than in conventional divalent  (and even monovalent) copper 
compounds. The decreasing exchange splitting is likely an additional 
factor affecting energy deepening of $3d$ states.

%**-----------------------------------------------------
The deepening of $d$ states across the series is accompanied by a
shift towards higher binding energies of the maxima of the Me $2p$
and Me $3s$ XPS spectra of Me[N(CN)$_2$]$_2$ in comparison with those
for $3d$ monoxides.\cite{Galakhov2, Galakhov}
This shift increases from about 0 eV for Mn[N(CN)$_2$]$_2$  to
about 3 eV for Cu[N(CN)$_2$]$_2$. Due to weak
metal-metal interactions in molecular crystals, especially in
Cu[N(CN)$_2$]$_2$ where $3d$ states are strongly localized, 
the screening of the final states is weaker than in the other cases, resulting
in larger binding energies of the valence bands and the core levels.
%**--------------------------------------------------

At the top of the valence band the detailed electronic structure
is determined by the filling of the transition metal $3d$ states. 
The local environment around Me sites is a distorted octahedron of N sites, 
and the ligand field splits the $3d$ states by approximately 1.5 eV into 
$t_{2g}$ and $e_g$ manifolds. In Mn[N(CN)$_2$]$_2$, with an exchange splitting 
of over 3 eV, the majority $d$ bands are fully occupied and there is a 
significant gap to the unoccupied minority $d$ bands.\cite{Pederson} 
As the number of $d$ electrons increases, the ligand-field splitting stays 
about the same while the exchange splitting decreases. The majority bands 
nevertheless remain full. The minority $t_{2g}$ bands are narrow and split 
by the tetragonal distortion of the local octahedral environment and 
by the longer-range crystal field. In the FM configuration, intersublattice 
$d-d$ interactions broaden the majority $t_{2g}$ bands, resulting in a 
two-fold subband of $d_{xz,yz}$ states with a gap to the $d_{xy}$ subband. 
(Axes are referenced to the local octahedral environment.) 
In the AF configuration, intersublattice $d-d$ interactions are suppressed, 
which narrows the $t_{2g}$ bands.  For Me = Fe, 
the ordering of the one-fold and two-fold subbands reverses in the FM phase. 
Thus, with one minority $t_{2g}$ electron 
per Fe site, AF Fe[N(CN)$_2$]$_2$ is metallic while FM Fe[N(CN)$_2$]$_2$ 
is insulating, as shown in Fig. \ref{dos_Fe}.  The lack of a gap in 
AF Fe[N(CN)$_2$]$_2$ with collinear spins contributes to its instability
with respect to the insulating FM phase.  In FM Co[N(CN)$_2$]$_2$, the two-fold 
subband fills, and in FM Ni[N(CN)$_2$]$_2$, the $t_{2g}$ manifold is 
completely occupied. The Jahn-Teller effect significantly enhances the 
tetragonal distortion of the octahedral environment around Cu sites in 
Cu[N(CN)$_2$]$_2$,\cite{Kurmoo, Batten} which increases the splitting 
between $e_g$ bands. 
The calculated GGA band gaps are listed in Table \ref{summary}, along with 
estimated gaps from diffuse reflectance spectroscopy.
The agreement is better for the compounds 
where $t_{2g}$ band is either completely empty or completely full.

The narrow width of some of the $d$ bands (Fig. \ref{dos_Fe})  suggests that electron 
correlations could be important in these materials.   Additional
evidence for this is the GGA's large underestimation of 
the band gaps in some of the compounds, particularly Co[N(CN)$_2$]$_2$. 
Furthermore, the fact that the measured magnetic moments in Fe[N(CN)$_2$]$_2$
and Ni[N(CN)$_2$]$_2$ are larger than the maximum
moments that could be created by spin alone
indicates the importance of the orbital moment, and 
therefore methods incorporating orbital dependence of the potential
could significantly improve the results. 
The situation is reminiscent of the monoxides of the $3d$ transition metals.  
As with the oxides, it is 
likely that methods beyond DFT-GGA are needed for a better description of 
the electronic and magnetic structure of the dicyanamides in the solid state.

\section{Conclusions}

We have carried out a theoretical and experimental investigation 
of the electronic structure of the  Me[N(CN)$_2$]$_2$ 
(Me = Mn, Fe, Co, Ni, Cu) molecular magnets. 
The general features of the valence bands are similar across
the series, with the primary exceptions being the location of the
$3d$ bands with respect to the Fermi level and the degree of hybridization
between Me $3d$ states and N $2p$ states. Going across the row from
Mn to Cu, the $d$ bands lie deeper in energy, and the
degree of hybridization with N $p$ states increases.  

Total-energy calculations indicate that the small differences in
structural parameters (for example, the angle of the superexchange
pathway)  found in these isostructural materials is not sufficient to 
account for differences in their magnetic ground states.    Rather,
the occupation of the $3d$ shell is likely a more significant factor.

For most of the materials in the series, our calculations underestimate 
either the band gap or the local magnetic moment.  
Together with the existence of narrow d-bands in these materials,
this suggests that a better description of the magnetic and electronic
structure of these materials will require a more accurate treatment 
of electron-electron interactions. 

\begin{acknowledgments}  
The authors thank M. R. Pederson for helpful discussions and for
communicating unpublished results.
This work was supported by the Research Council of the President of the Russian 
Federation under Grant NSH-1026.2003.2, the Russian Foundation for Basic 
Research under Project 02-02-16674, the  President's NSERC fund of the 
University of Saskatchewan, Deutsche Forschungsgemeinschaft (DFG) under
the priority programme SPP 1137 Molekularer Magnetismus, 
the US National Science Foundation under 
Grant DMR-0210717, and the US Office of Naval Research under 
Grant N00014-02-1-1046. 
Work at the Advanced Light Source at Lawrence Berkeley National Laboratory 
was supported by the U.S. Department of Energy under 
Contract No. DE-AC03-76SF00098. 
Work at the Advanced Photon Source is supported by the U.S. Department of Energy, 
Office of Science, Office of Basic Energy Sciences under Contract No. W-31-109-ENG-38.
\end{acknowledgments}

\newpage
\begin{figure}
\caption{Me $3s$ x-ray photoelectron spectra of Me[N(CN)$_2$]$_2$ (points)
and simple oxides MeO (thick lines). The maxima of the Me $3s$ spectra of
monoxides have been shifted in binding energy to facilitate comparison with the 
Me[N(CN)$_2$]$_2$ spectra. For Me[N(CN)$_2$]$_2$, fitting curves are shown with 
thin lines. 
\label{XPS_3s}}
\end{figure}
\begin{figure}
\caption{Non-resonant (a) carbon and (b) nitrogen $K_\alpha$ XES of 
Me[N(CN)$_2$]$_2$ compounds. The N $K_\alpha$ XES for the reference compound 
Na$_4$Fe(CN)$_6$(OH$_2$O) is included for comparison. 
\label{XES}}
\end{figure}
\begin{figure}[t]
\caption{Resonant N $K_\alpha$ XES of Mn[N(CN)$_2$]$_2$.  
Spectra obtained with different excitation energies near the N $1s$ 
threshold are shown in (a). The non-resonant spectrum $d$ is compared to superpositions
of resonant spectra $a$, $b$, and $c$ in (b).  The non-resonant spectrum is
better represented by a superposition of $b$ and $c$ than by 
a superposition of $a$ and $c$.
\label{RXES}}
\end{figure}
\begin{figure}[ht]
\caption{Comparison of XPS VB and XES of constituents for Mn[N(CN)$_2$]$_2$
in the binding energy scale.  \hfill
\label{Mn_XPS} }
\end{figure}
\begin{figure}
\caption{Comparison of XPS VB and XES of constituents for Fe[N(CN)$_2$]$_2$
in the binding energy scale.  \hfill
\label{Fe_XPS}}
\end{figure}
\begin{figure}[ht]
\caption{Comparison of XPS VB and XES of constituents for Co[N(CN)$_2$]$_2$
in the binding energy scale.   \hfill
\label{Co_XPS} }
\end{figure}
%---
\begin{figure}
\caption{Comparison of XPS VB and XES of constituents for Ni[N(CN)$_2$]$_2$
in the binding energy scale.  \hfill
\label{Ni_XPS} }
\end{figure}
\begin{figure}
\caption{Comparison of XPS VB and XES of constituents for Cu[N(CN)$_2$]$_2$
in the binding energy scale.  \hfill
\label{Cu_XPS} }
\end{figure}
\begin{figure}
\caption{Total and partial density of states calculated for ferromagnetic 
Ni[N(CN)$_2$]$_2$. The total DOS in the bottom panel has been 
broadened by 1.0 eV, while the site-projected  DOS in the
top three panels have been broadened by 0.2 eV.  
\label{dos_Ni}}
\end{figure}
\begin{figure}
\caption{Cu and Ni $d$-densities of states calculated for FM Cu[N(CN)$_2$]$_2$ and 
FM Ni[N(CN)$_2$]$_2$. 
To emphasize gross features in the DOS, Methfessel-Paxton (fifth order) smearing of 1 eV is used. 
The band gaps are masked by this broadening, and the Fermi level is set to 0 eV. 
The inset shows total FM Cu[N(CN)$_2$]$_2$ densities of states calculated with the linear tetrahedron method. 
\label{dos_Cu}}
\end{figure}
\begin{figure}[ht]
\caption{Fe $d$-density of states calculated for Fe[N(CN)$_2$]$_2$
with (a) AF spin ordering and (b) FM spin ordering. The linear
tetrahedron method is used to calculate the DOS. 
\label{dos_Fe}}
\end{figure}

%\newpage
\clearpage
\begin{table*}
\caption{Calculated total-energy differences between different spin
configurations (in eV/Me site),  local spin magnetic moments (in $\mu_B$), and
band gaps (in eV).  
For Me = Mn and Fe, the calculated spin moments and gaps are given for the antiferromagnetic state,
while for Me = Co, Ni, and Cu, results for the ferromagnetic state are presented. 
Experimental moments from neutron powder
diffraction experiments reported in Refs. \onlinecite{Kmety1,Kmety2,Kmety0} 
are shown for comparison and include both spin and orbital contributions.  
Estimates for band gaps obtained from diffuse reflectance 
spectroscopy on powder samples at room temperature are also included.
\label{summary}}
 \begin{ruledtabular}
 \begin{tabular}{ccccccc}
  & \multicolumn{2}{c}{$\Delta E$} & \multicolumn{2}{c}{$\mu_{Me}$}  &
    \multicolumn{2}{c}{$E_{g}$} \\
  & NM-AF & AF-FM & calc  & expt & calc & expt \\
\hline
Mn[N(CN)$_2$]$_2$ &2.490 &-0.018 & 4.50 &   4.61 & 2.12     & 2.07 \\
Fe[N(CN)$_2$]$_2$ &0.862 & 0.025 & 3.59 &   4.53 & 0.00\footnote{Calculations yield metallic state for AF Fe[N(CN)$_2$]$_2$ (see text).}    & 1.46 \\
Co[N(CN)$_2$]$_2$ &0.735 & 0.023 & 2.54 &   2.67 & 0.24     & 2.40 \\
Ni[N(CN)$_2$]$_2$ &0.939 & 0.015 & 1.59 &   2.21 & 1.36     & 1.97 \\
Cu[N(CN)$_2$]$_2$ &0.287 & 0.000 & 0.65 &   1.05 & 0.52     & 1.91
\end{tabular}
\end{ruledtabular}
\end{table*}

\begin{table*}
\caption{XPS core level binding energies and location of $3d$-band maxima (in eV) of Me[N(CN)$_2$]$_2$, with
Me = Mn, Fe, Co, Ni, Cu.
\label{bindingenergy}}
\begin{ruledtabular}
\begin{tabular}{lccccc}
   & Mn[N(CN)$_2$]$_2$ & Fe[N(CN)$_2$]$_2$ & Co[N(CN)$_2$]$_2$ & Ni[N(CN)$_2$]$_2$ & Cu[N(CN)$_2$]$_2$ \\
\hline
C $1s$ & 287.7 & 287.7 & 287.1 & 288.0 & 288.0 \\
N $1s$ & 399.6; 398.0 & 399.6; 398.3  & 399.6; 398.5 & 399.6; 397.2 & 399.6; 397.8 \\
Me $2p_{3/2}$ & 642.4 & 710.1 & 782.2 & 857.2 & 936.5 \\
Me  $3d$ & 4.0 & 4.1 & 4.2 & 4.7 & 6.0 \\
\end{tabular}
\end{ruledtabular}
\end{table*}

\newpage

\begin{figure*}
\includegraphics[width=465pt,height=665pt]{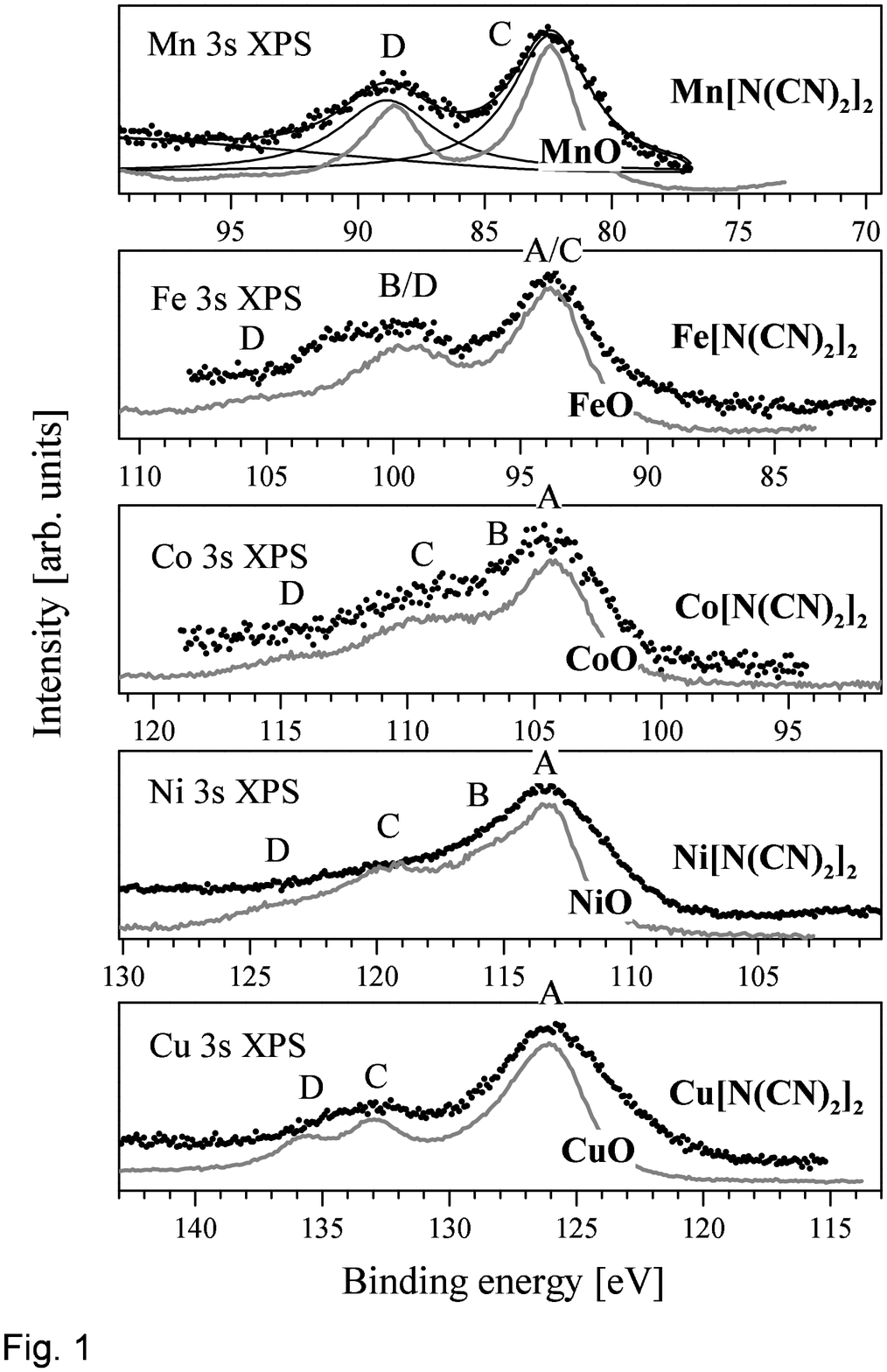}
\end{figure*}

\newpage
\begin{figure*}
\includegraphics[width=465pt,height=665pt]{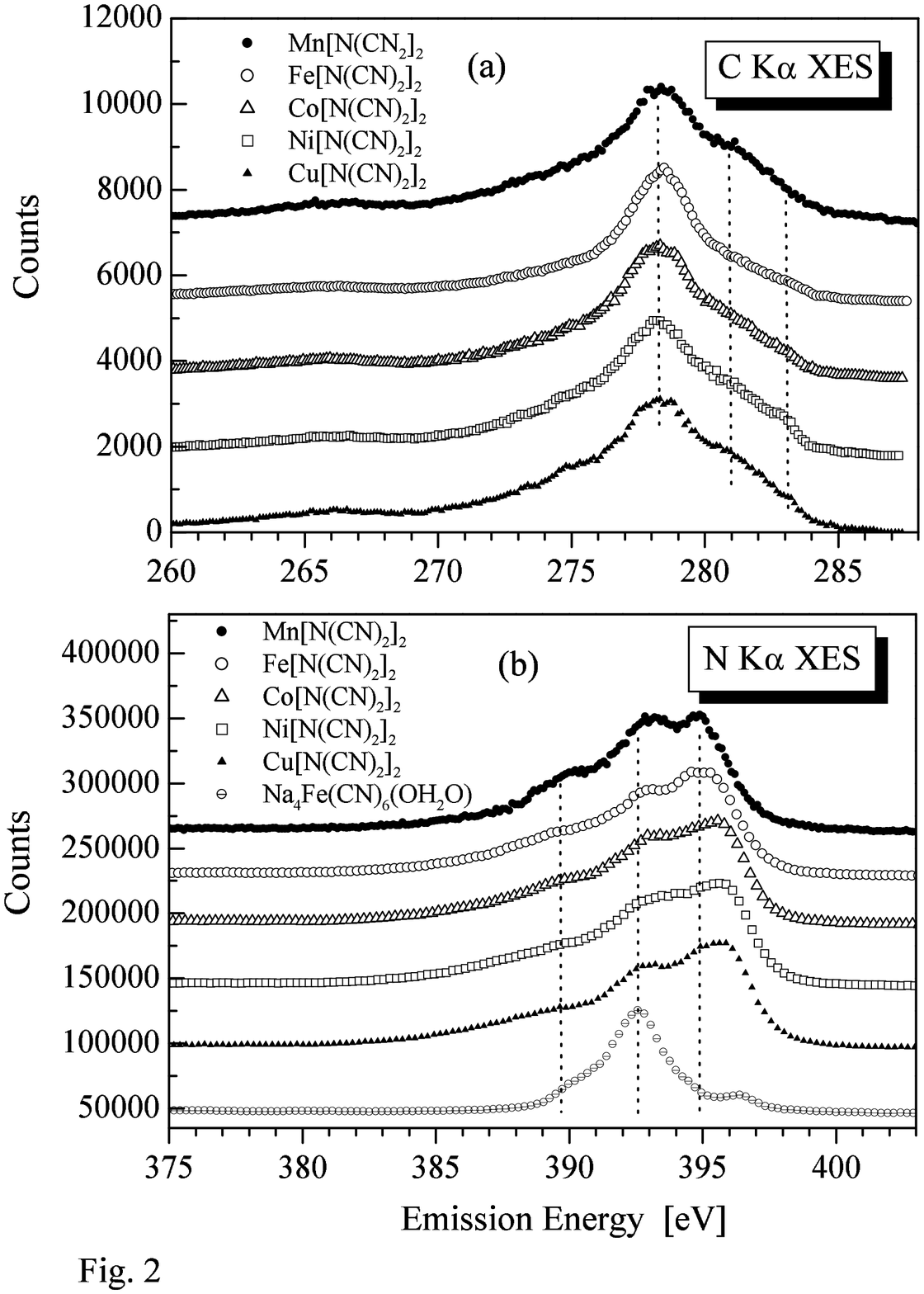}
\end{figure*}

\newpage
\begin{figure*}
\includegraphics[width=465pt,height=665pt]{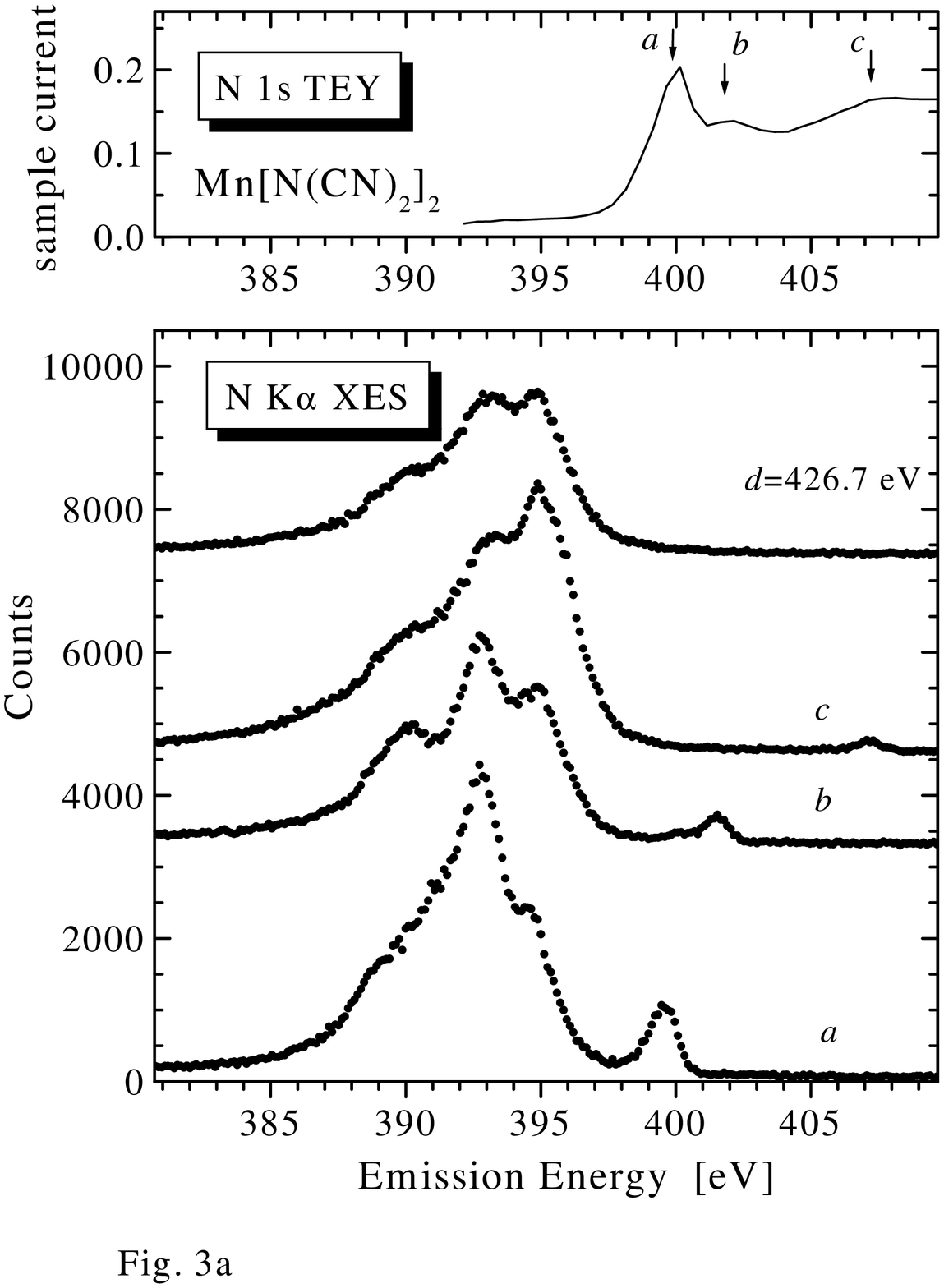}
\end{figure*}

\newpage
\begin{figure*}
\includegraphics[width=465pt,height=665pt]{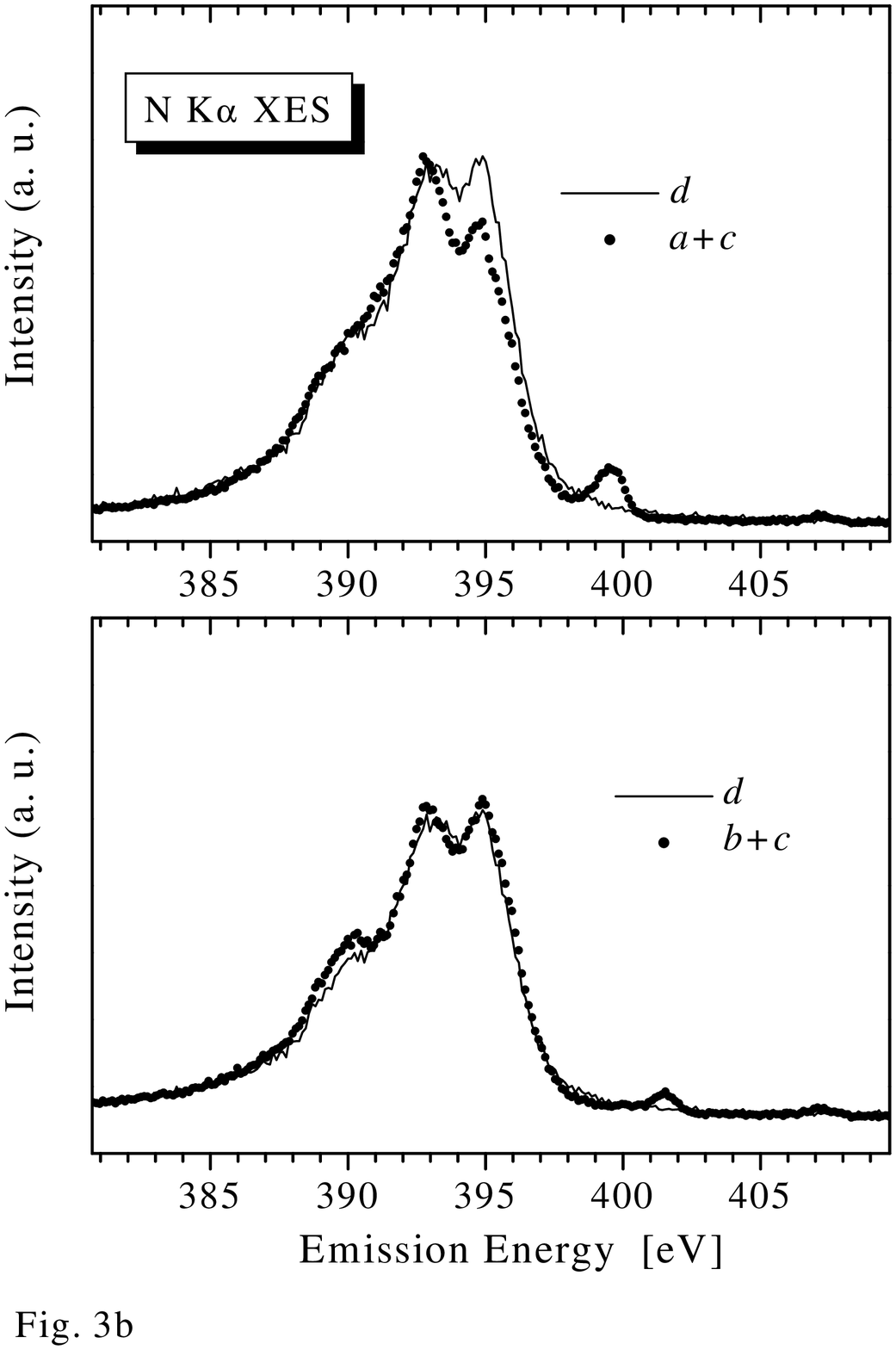}
\end{figure*}

\newpage
\begin{figure*}
\includegraphics[width=465pt,height=665pt]{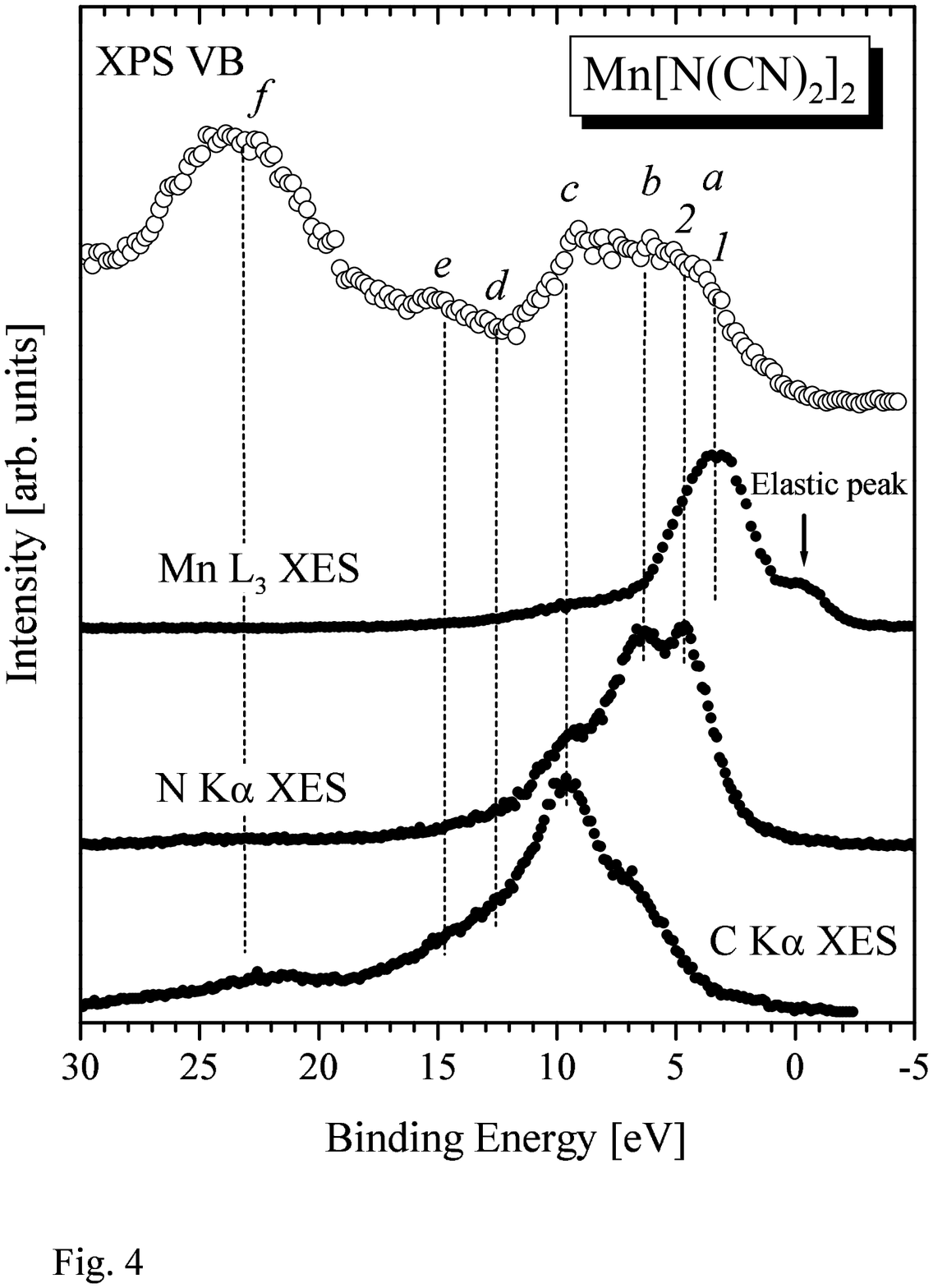}
\end{figure*}

\newpage
\begin{figure*}
\includegraphics[width=465pt,height=665pt]{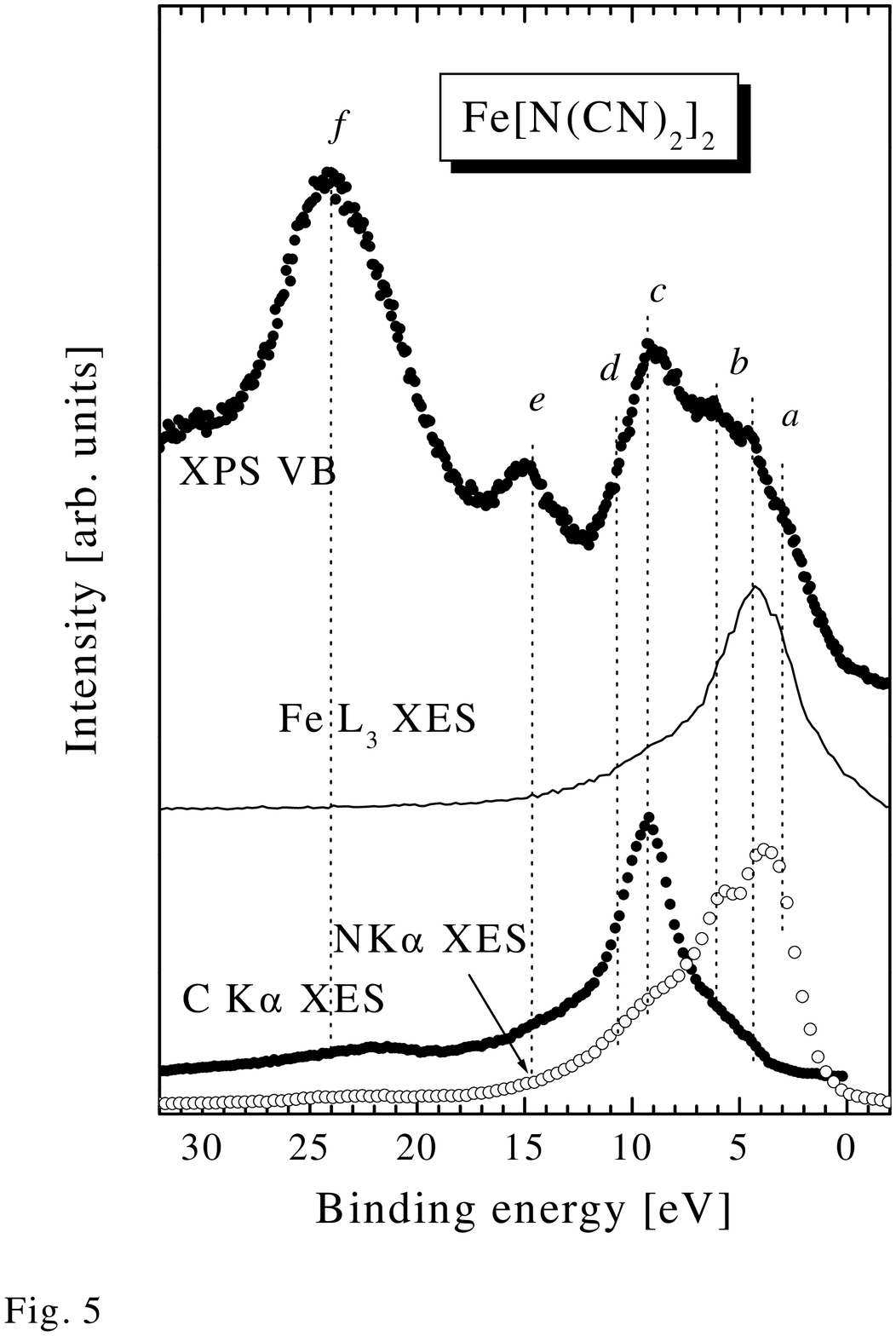}
\end{figure*}

\newpage
\begin{figure*}
\includegraphics[width=465pt,height=665pt]{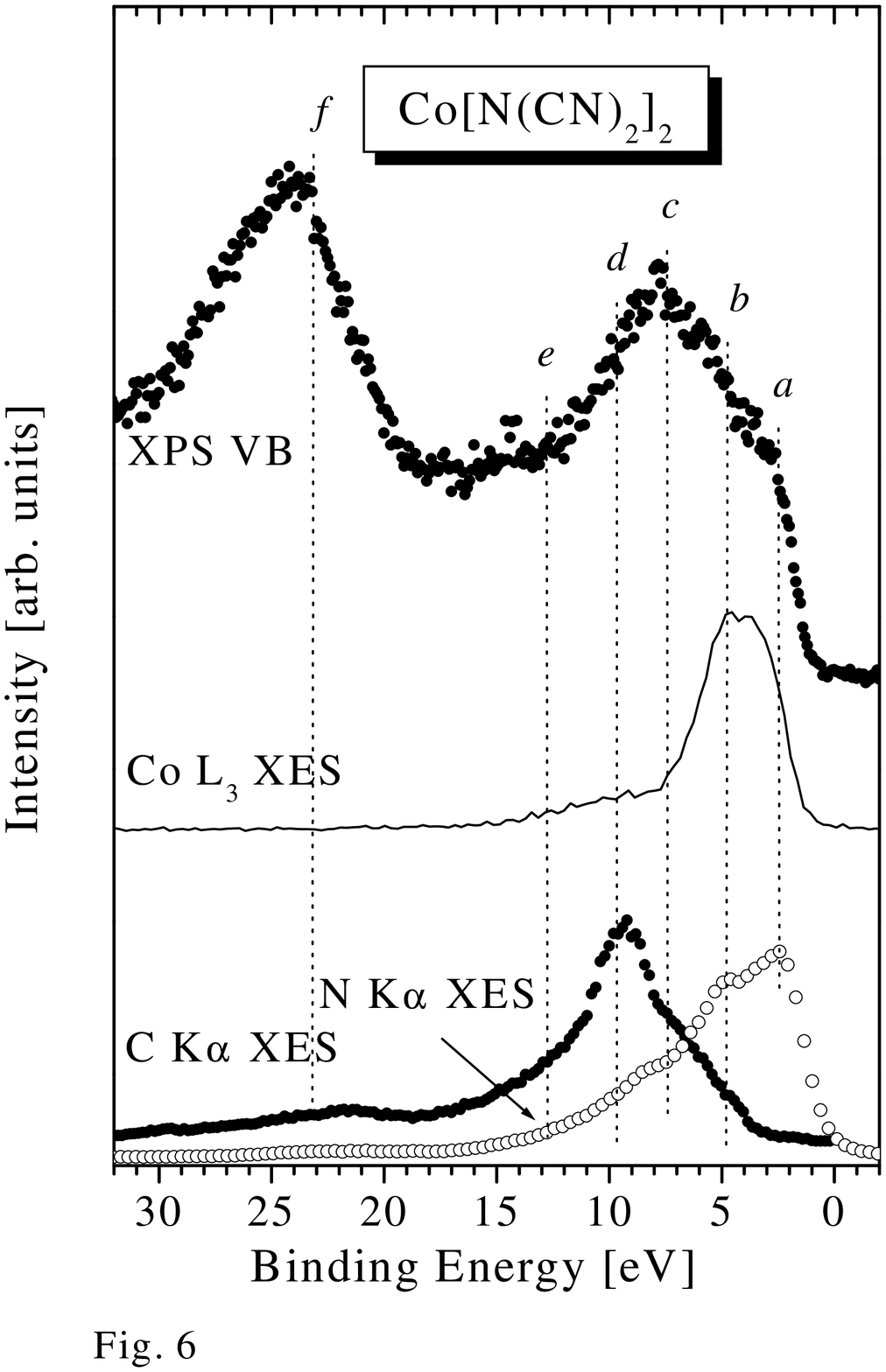}
\end{figure*}

\newpage
\begin{figure*}
\includegraphics[width=465pt,height=665pt]{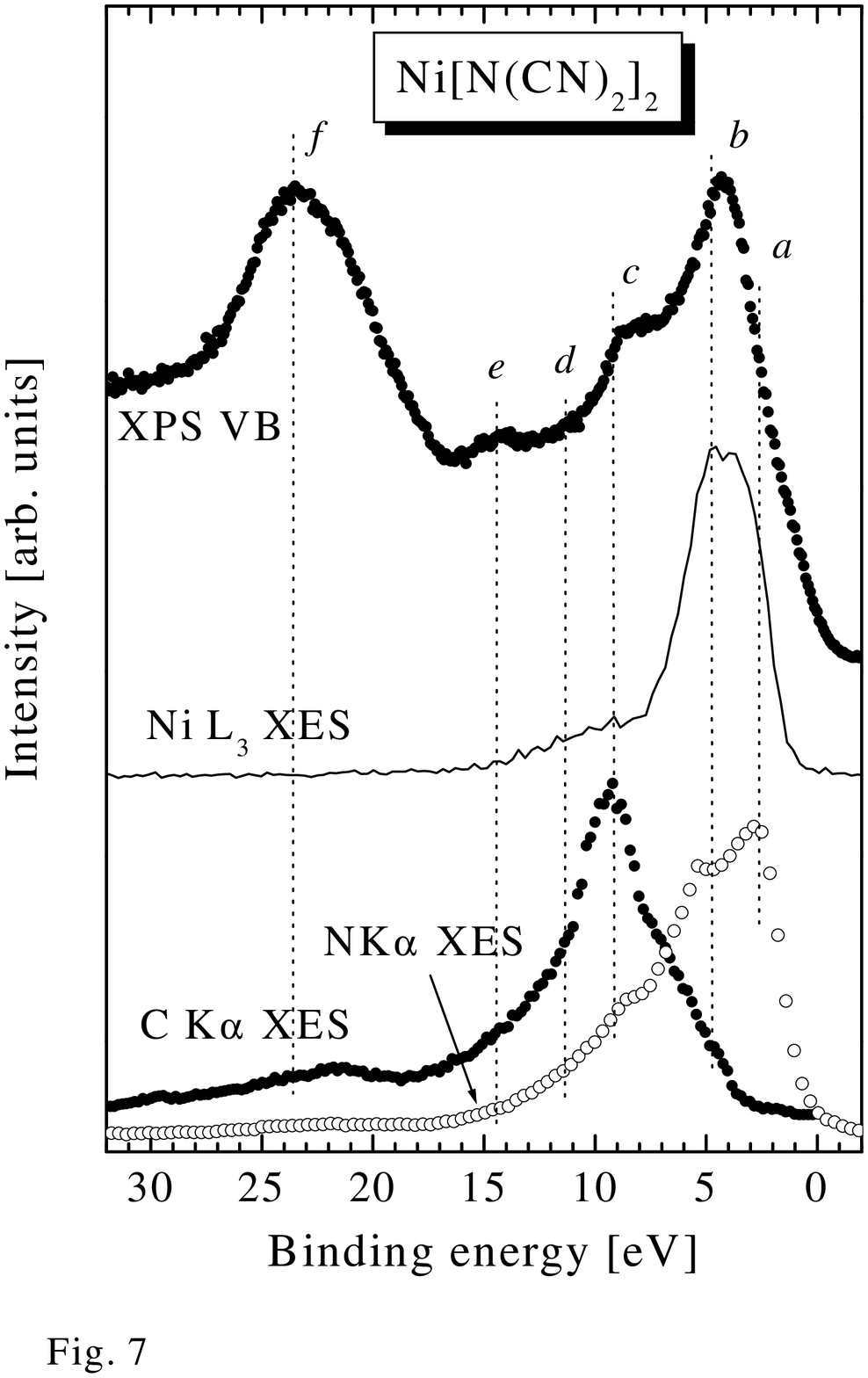}
\end{figure*}

\newpage
\begin{figure*}
\includegraphics[width=465pt,height=665pt]{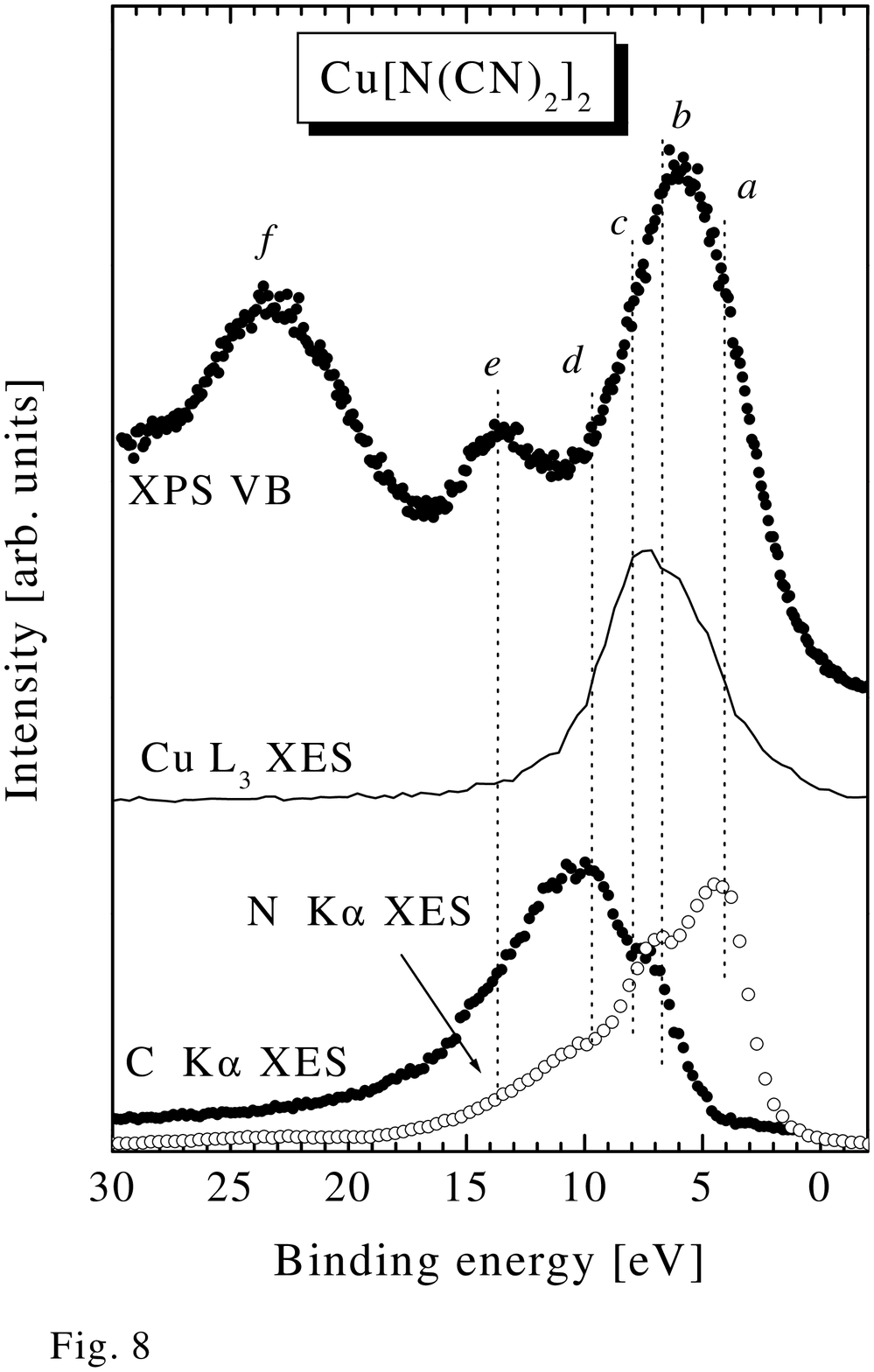}
\end{figure*}

\newpage
\begin{figure*}
\rotatebox[]{270}{\includegraphics[width=365pt,height=465pt]{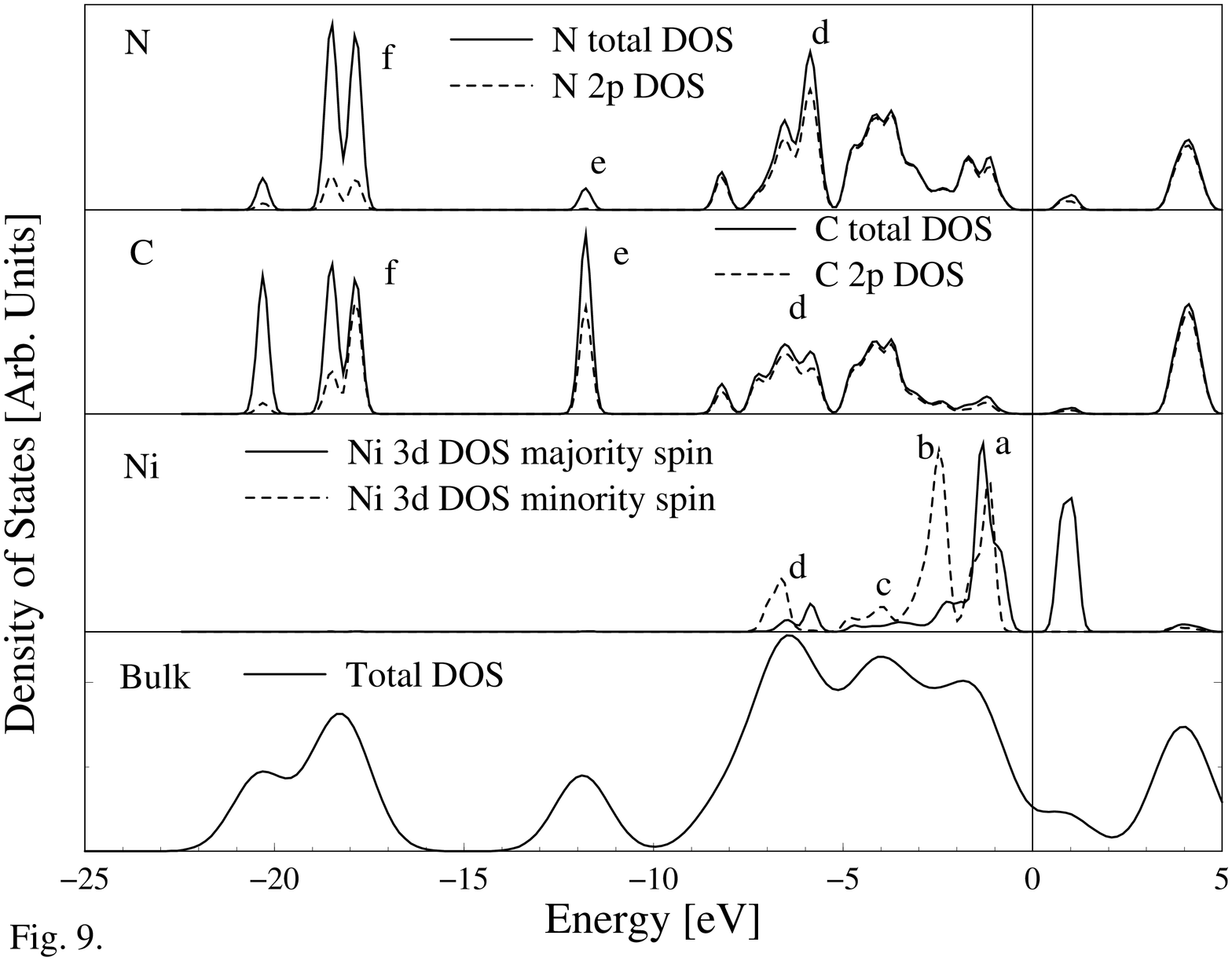}}
%\rotatebox[]{270}{\includegraphics[width=400pt,height=465pt]{Fig9.pdf}}
\end{figure*}

\newpage
\begin{figure*}
\rotatebox[]{270}{\includegraphics[width=365pt,height=465pt]{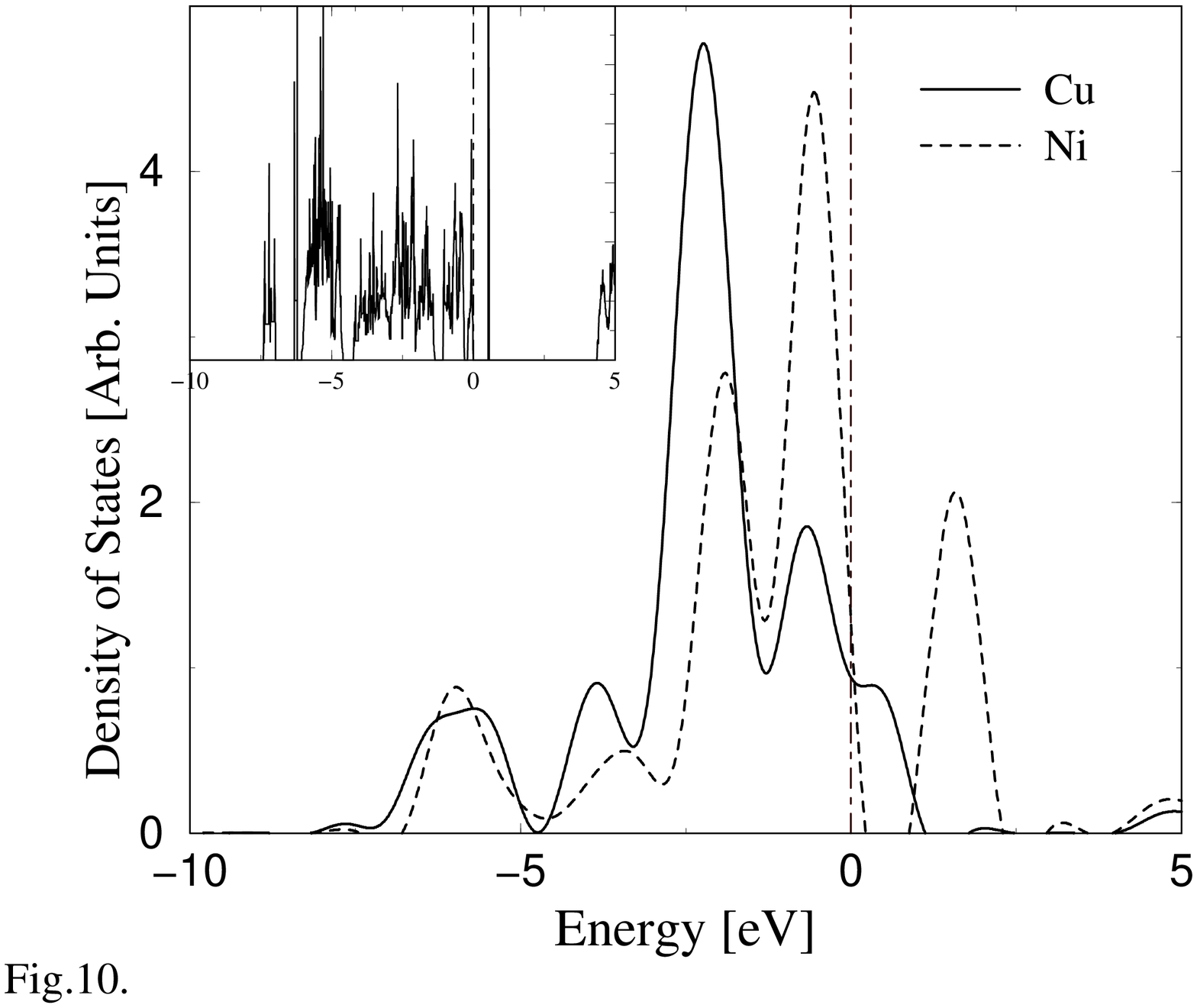}}
%\rotatebox[]{270}{\includegraphics[width=400pt,height=465pt]{Fig10.pdf}}
\end{figure*}

\newpage
\begin{figure*}
\rotatebox[]{270}{\includegraphics[width=365pt,height=465pt]{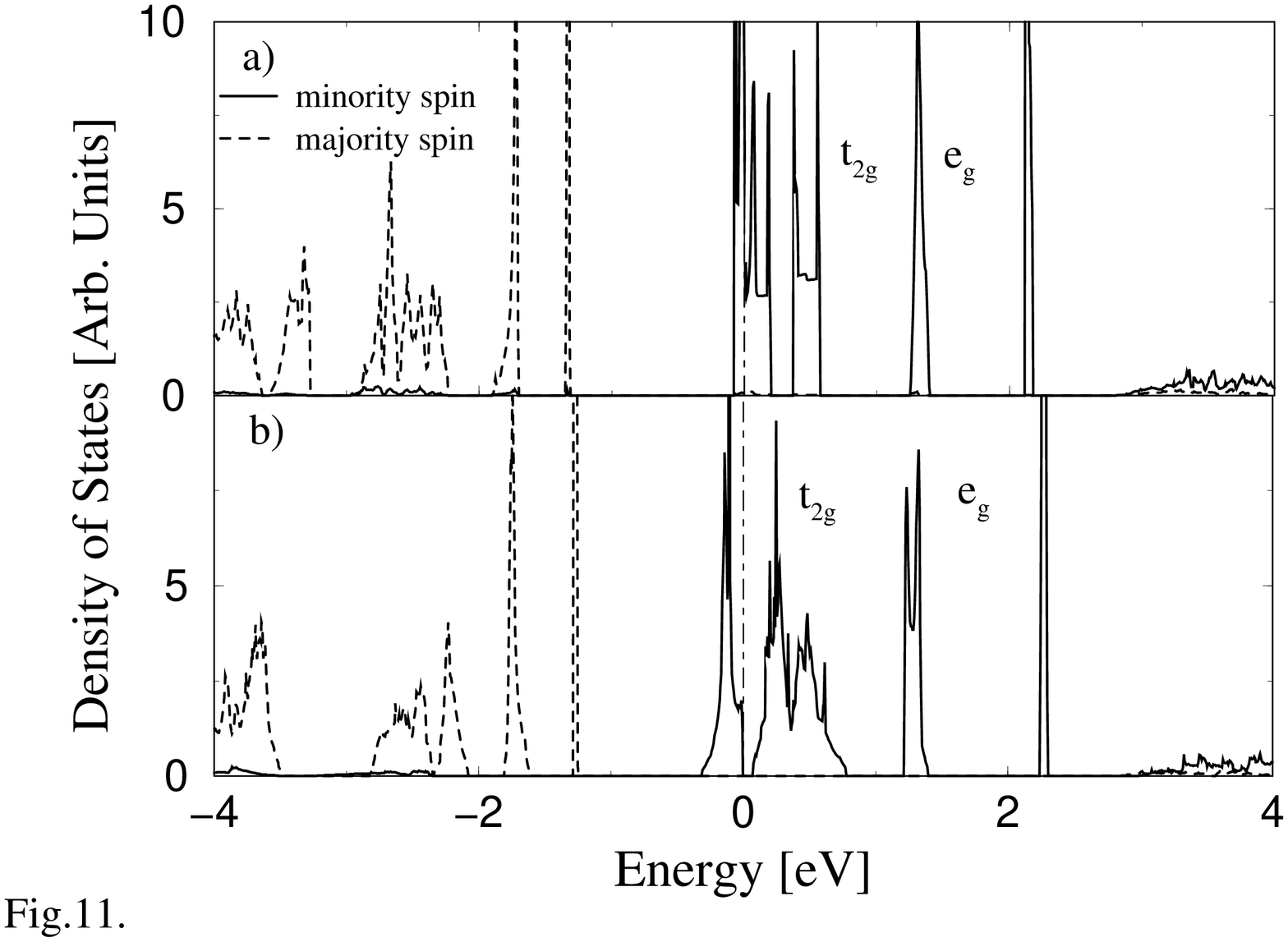}}
%\rotatebox[]{270}{\includegraphics[width=400pt,height=465pt]{Fig11.pdf}}
\end{figure*}


\begin{thebibliography}{99}
\bibitem{Manriquez} J. M. Manriquez, G. T. Yee, R. S. McLean,
A. J. Epstein, and J. S. Miller, Science {\bf 252}, 1415 (1991).
\bibitem{Ferlay} S. Ferlay, T. Mallah, R. Ouahes, P. Veillet, and M.
Verdaguer, Nature {\bf 378}, 701 (1995).
\bibitem{Kmety3} J. L. Manson, C. R. Kmety, Q. Huang, J. W. Lynn,
G. M. Bendele, S. Pagola, P. W. Stephens, L. M. Liable-Sands, A. L. 
Rheingold, A. J. Epstein, and J. S. Miller, Chem. Mater. {\bf 10}, 2552
(1998).
\bibitem{Batten} {S. R. Batten, P. Jensen, B. Moubaraki, K. S. Murray, 
and R. Robson}, {Chem. Commun.} {\bf 3}, 439 (1998).
\bibitem{Kurmoo}{M. Kurmoo and C. J. Kepert}, {New J. Chem.} {\bf 22}, 
1515 (1998).
\bibitem{Kmety1}{C. R. Kmety, Q. Huang, J. W. Lynn, R. W. Erwin, J. L. Manson, 
S. McCall, J. E. Crow, K. L. Stevenson, J. S. Miller, and A. J. Epstein}, 
{Phys. Rev. B} {\bf 62},  5576  (2000).
\bibitem{Kmety2} C. R. Kmety, J. L. Manson,
Q. Huang, J. W. Lynn, R. W. Erwin, J. S. Miller, A. J. Epstein,
{Phys. Rev. B} {\bf 60}, 60 (1999).
\bibitem{thesis}{C. R. Kmety}, Ph.D. Thesis, The Ohio State University, 2000.
\bibitem{Pederson} {M. R. Pederson, A. Y. Liu, T. Baruah, E. Z. Kurmaev, 
A. Moewes, S. Chiuzb\v{a}ian, M. Neumann, C. R. Kmety, K. L. Stevenson, 
D. Ederer}, {Phys. Rev. B} {\bf 66}, 014446 (2002).
\bibitem{Jia} J. J. Jia, T. A. Callcott, J. Yurkas, A. W. Ellis, 
F. J.  Himpsel,  M. G. Samant,  J. St\"{o}hr, D. L. Ederer, J. A. Carlisle, 
E. A. Hudson, L. J. Terminello,  D. K. Shuh, and 
R. C. C. Perera,  Rev. Sci. Instrum. {\bf 66}, 1394 (1995).
\bibitem{Moulder} G. Beamson, D. Briggs, { \it High
Resolution XPS of Organic Polymers: the Scienta ESCA300 Database} (Wiley, Chichester, 1992).
\bibitem{vasp}{G. Kresse and J. Furthm\"uller}, 
{Comput. Mater. Sci.} {\bf 6},  15  (1996); 
{Phys. Rev. B} {\bf 54},  11169  (1996).
\bibitem{GGA1}{Y. Wang and J. P. Perdew}, {Phys. Rev. B} {\bf 44},  
13298  (1991).
\bibitem{GGA2}{J. P. Perdew, K. Burke, and M. Ernzerhof}, 
{Phys. Rev. Lett.} {\bf 77}, 3865 (1996).
\bibitem{Monkhorst-Pack} H. J. Monkhorst and J. D. Pack, Phys. Rev. B 
{\bf 13}, 5188 (1976). 
\bibitem{Pederson1} M. R. Pederson and T. Baruah (private communication).
\bibitem{Ruiz} E. Ruiz, J. Solid State Chem. (to be published).
\bibitem{Anderson} P. W. Anderson, Phys. Rev. {\bf 79}, 350 (1950); Phys. Rev. {\bf 115}, 2 (1959).
\bibitem{Goodenough} J. B. Goodenough, {\sl Magnetism and the Chemical Bond} (Wiley, New York, 1963).
\bibitem{Gudel} H. Weihe, H. U. G\"udel, H. Toftlund, Inorg. Chem. {\bf 39}, 1351 (2000).
\bibitem{Figgis} B. N. Figgis, {\it Introduction to Ligand Fields}
(Wiley, New York, 1966).
\bibitem{VanVleck} J. H. Van Vleck, Phys. Rev. B {\bf 45}, 405 (1934). 
\bibitem{Galakhov2} V. R. Galakhov, M. Demeter, S. Bartkowski,
M. Neumann, N. A. Ovechkina, E. Z. Kurmaev, N. I. Lobachevskaya, Ya. M. Mukowskii, 
J. Mitchell, D. L. Ederer, Phys. Rev. B {\bf 65}, 113102 (2002).
\bibitem{Galakhov} V. R. Galakhov, L. D. Finkelstein, D. A. Zatsepin, 
E. Z. Kurmaev, A. A. Samokhvalov, S. V. Naumov, G. K. Tatarinova, M. Demeter, 
S. Bartkowski, M. Neumann, and A. Moewes, Phys. Rev. B {\bf 62}, 
4922 (2000).
\bibitem{Kmety0} C. R. Kmety, J. L. Manson, S. McCall, J. E. Crow,
K. L. Stevenson, and A. J. Epstein, J. Magn. Magn. Mater. {\bf 248}, 52 (2002).
\end{thebibliography}
\end{document}